\documentclass[conference]{IEEEtran}
\IEEEoverridecommandlockouts
\usepackage{cite}
\usepackage{amsmath,amssymb,amsfonts}
\usepackage{algorithmic}
\usepackage{graphicx}
\usepackage{textcomp}
\usepackage{tabularx, multirow, booktabs}
\usepackage{xcolor}
\def\BibTeX{{\rm B\kern-.05em{\sc i\kern-.025em b}\kern-.08em
    T\kern-.1667em\lower.7ex\hbox{E}\kern-.125emX}}
\begin{document}

\title{Strategy Phasing of Cyber Attacks on
Digital Substations
}

\author{\IEEEauthorblockN{Akila Herath, Chen-Ching Liu}
\IEEEauthorblockA{\textit{Department of Electrical and Computer Engineering
} \\
\textit{Virginia Tech}\\
Blacksburg, VA, USA \\
akilaasansana@vt.edu, ccliu@vt.edu}
\and
\IEEEauthorblockN{Junho Hong, Kuchan Park}
\IEEEauthorblockA{\textit{Department of Electrical and Computer Engineering
} \\
\textit{University of Michigan-Dearborn}\\
Dearborn, MI, USA \\
jhwr@umich.edu, kuchan@umich.edu}
}

\maketitle

\begin{abstract}
Digital substations that comply with IEC 61850 have improved the operational efficiency of modern power systems. However, adversaries can abuse IEC 61850 communication to manipulate circuit breaker operations in substations, which can result in severe system impacts. These cyber attacks are crafted based on broader multi-phase strategies. The existing intrusion detection systems (IDSs) often flag only isolated symptoms. Thus, there is a lack of context in the attack phase to support the deployment of mitigation measures. This paper proposes Substation Cyber Attack Strategy Phasing (SubCASP), a Hidden Markov Model(HMM)- based method that fuses IDS data logs to infer the current attack phase, next attack phase, and retrospective attack path. The attack phases are derived from an ATT\&CK-based threat modeling. The SubCASP model is trained and evaluated on a reproducible attack-graph dataset. Test results are presented to demonstrate the robustness of SubCASP for various IDS observability levels and missing IDS data logs scenarios.
\end{abstract}

\section{Introduction}
The transition of conventional substations to digital substations with IEC 61850 creates a tightly coupled cyber-physical system for protection, control, and monitoring. This modernization improves interoperability and life-cycle costs, yet it also expands the surface for cyber attacks. The critical issue here is that the compromise of cyber components can quickly propagate to trigger physical system misoperations. 

IEC 61850 defines three communication protocols. Manufacturing Message Specification (MMS) is based on a client–server configuration that supports reporting/ control commands. Generic Object Oriented Substation Event (GOOSE) and Sampled Values (SV) are operated as multicast protocols. GOOSE is transmitted for event-driven control/status messages. Time-critical analog measurements from the physical layer are streamed as SV. Adversaries can abuse each of these protocols: spoofed MMS control commands; forged GOOSE to issue unauthorized trips/closes; SV manipulation to trigger protection logics; or denial-of-service (DoS) attack to suppress legitimate traffic. The key consequence is the disruption of circuit-breaker (CB) operations, causing possible equipment damage and system instability.

In practice, cyber attacks targeting CB operations rarely follow a single step. Successful intrusions typically reflect a broader, multi-phase strategy. Cyber-threat modeling frameworks such as MITRE ATT\&CK \cite{mitre}, STRIDE, and Kill Chain \cite{kill} offer a vocabulary and structure for these attack phases and prerequisites. Modeling these attack phases clarifies which progressions are feasible, what evidence each phase emits, and where mitigation can interdict.

To aid in defense against digital substation cyber attacks, numerous detection mechanisms have been proposed. These include MMS command validation schemes, authentication per IEC 62351-6 with sequence number checks for SV and GOOSE \cite{iec62351}, machine-learning-based detectors \cite{mlsdn}, and host-based defense techniques \cite{ids}. However, these methods typically isolate single attack events and return an alert without pinpointing them within the overall strategy. The key limitations are:
\begin{itemize}
\item No attack phase context: Alerts do not infer the adversary’s current state of the multi-phase attack.
\item No objective context: Alerts do not indicate the adversary’s end goal, limiting risk assessment.
\item No path awareness: Detection mechanisms cannot reconstruct the phases that the adversary's strategy followed.
\end{itemize}
If the above limitations are addressed by utilizing the evidence from detection mechanisms, mitigation can be deployed more accurately along the attack chain.

This paper addresses this research gap with a strategy-phasing approach. The contributions of the proposed methodology are as:
\begin{enumerate}
\item Graph-constrained Hidden Markov Model (HMM) that performs substation cyber attack strategy-phasing (SubCASP) from heterogeneous IDS observations.
\item Cyber threat modeling-based attack graph specialized to digital substations pursuing CB operation disruption
\item An evaluation using a reproducible, attack-graph dataset to test the proposed SubCASP method across IDS observability levels and missing IDS log levels.
\end{enumerate}

The remainder of the paper is structured as follows: Section
II defines cyber attack strategy in a digital substation. Section III elaborates on the cyber threats modeling. Section IV focuses on
the formulation of the HMM application for SubCASP, and Section V introduces the reproducible attack-graph dataset. Section VI presents numerical results. Section VII concludes with recommendations for future work.

\section{Cyber Attack Strategy in a Digital Substation}

The architecture of a digital substation is illustrated in Fig. 1. Measurements from the physical layer are digitized by Merging Units (MUs) and published as SV on the process bus to subscribing Intelligent Electronic Devices (IEDs). IEDs exchange fast event/control GOOSE messages, multicasted on the process bus to actuate CB operations via the MU. For supervisory and engineering functions, IEDs use MMS at the station bus to report status/measurements and to accept configuration and controls. Station-level devices, Human Machine Interface (HMI), and Remote Terminal Unit (RTU) send operator or control-center actions to IEDs over MMS.

A primary goal of an adversary attacking a substation is to compromise IEC 61850-based communications and disrupt the physical layer operation. Such an adversary has a strategy to progress through a set of attack phases on different devices of the substation. A phase of the attack is a milestone that adversaries aim to achieve during an attack as part of the broader strategy. An example of such a strategy is demonstrated in Fig. 1. An adversary initially accesses the substation local area network (LAN) by exploiting the RTU at the station level, discovers other assets, and obtains access to an IED at the process level. Then the adversary changes the IED configurations to send a malicious CB operation command to MU. Finally, the CB operation is implemented through MU.

\begin{figure}[!t]
\centering
\includegraphics[width=0.40\textwidth]{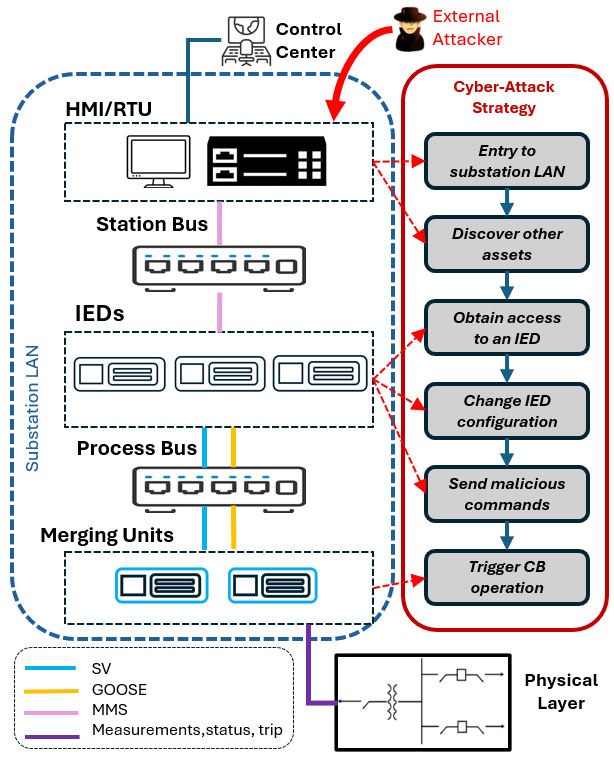}
\caption{An illustration of a cyber attack strategy in a digital substation.}
\label{fig_1}
\end{figure}

\section{Cyber Threats Modeling and Attack Graph}

The MITRE ATT\&CK framework for Industrial Communication Systems (ICSs) is used to define the attack phases in an adversary strategy to disrupt CB operations. Compared with similar cyber threat modeling tools, ATT\&CK provides a detailed representation of adversary actions, including their interrelationships, impact on adversarial goals, connections to mitigation techniques and data sources, and targeting of specific platforms. 

The key cyber attack phases identified with the ATT\&CK framework, with their definitions for this study, are given in Table 1. External adversaries may access the substation LAN by gaining the initial entry to the RTU through unauthorized access via the control center or by breaching remote access points. On-site adversaries may use a test set or an engineering laptop to connect to the HMI, station bus, or process bus \cite{mitre}. These initial entry acts are defined under \textit{'Initial Access'}.  

Depending on the entry point, the adversary may proceed through the intermediate phases, \textit{'Execution', 'Privilege Escalation', 'Discovery',} and \textit{'Lateral Movement'.} From an RTU/HMI foothold, an adversary can (i) issue MMS controls directly to target IEDs or (ii) pivot to the IEDs. On the station bus, the objective is to obtain MMS client/server privileges to send unauthorized configuration/command traffic. On the process bus, the goal is to inject malicious GOOSE/SV; because these are multicast and typically unauthenticated, an adversary with LAN access can often publish or interfere without additional privilege elevation.

If the adversary gains entry to an IED, they can alter the configuration or protection logic to manipulate circuit-breaker (CB) operation. Across paths, the campaign ultimately converges to \textit{'Inhibit Response Function'} or \textit{'Impair Process Control'} attack phases, as defined in ATT\&CK framework for manipulation of controllers such as IEDs and MUs. The ultimate phase, \textit{'Impact'}, is reached once the attack propagates to the hard-wired connection that links MU with the CB.  

\begin{table}[t]
\caption{Attack phases mapped to digital substation components}
\label{tab:attack-phases}
\centering
\footnotesize
\renewcommand{\arraystretch}{1.12}
\begin{tabularx}{\linewidth}{@{}|p{1.3cm} |p{2.6cm} |X|@{}}
\hline
\textbf{Device} & \textbf{Attack phase} & \textbf{Description} \\
\hline
\multirow{3}{*}{RTU/HMI}
  & Initial Access (IA) & Gain entry via HMI internally or externally breaching the remote access points. \\
  \cline{2-3}
  & Execution (Ex) & Execute scripts to perform unauthorized RTU operations. \\
  \cline{2-3}
  & Privilege Escalation (PE) & Obtain access to change configuration and send MMS commands. \\
\hline
\multirow{3}{*}{Station Bus}
  & Initial Access (IA) & Gain entry to a port internally. \\
  \cline{2-3}
  & Discovery (Dis) & Identify the hosts in MMS client/server communication. \\
  \cline{2-3}
  & Privilege Escalation (PE) & Obtain higher-privilege access to send MMS commands. \\
\hline
\multirow{5}{*}{IED}
  & Lateral Movement (LM) & Gain entry by extending access to RTU/HMI. \\
  \cline{2-3}
  & Execution (Ex) & Execute scripts to perform unauthorized IED operations. \\
  \cline{2-3}
  & Privilege Escalation (PE) & Obtain access to change configuration and send GOOSE commands. \\
  \cline{2-3}
  & Inhibit Response Function (IRF) & Prevent IED from sending GOOSE in response to events. \\
  \cline{2-3}
  & Impair Process Control (IPC) & Send unauthorized GOOSE commands to MU. \\
\hline
\multirow{2}{*}{Process Bus}
  & Initial Access (IA) & Gain entry to a port internally. \\
  \cline{2-3}
  & Discovery (Dis) & Capture GOOSE and SV multicast packets. \\
\hline
\multirow{3}{*}{MU}
  & Inhibit Response Function (IRF) & Prevent MU from receiving CB operation commands. \\
  \cline{2-3}
  & Impair Process Control (IPC) & Send unauthorized CB operation commands. \\
  \cline{2-3}
  & Impact (Imp) & Damage, interrupt, or disrupt physical layer operation. \\
\hline
\end{tabularx}
\vspace{-1mm}
\end{table}

\begin{figure}[!t]
\centering
\includegraphics[width=0.45\textwidth]{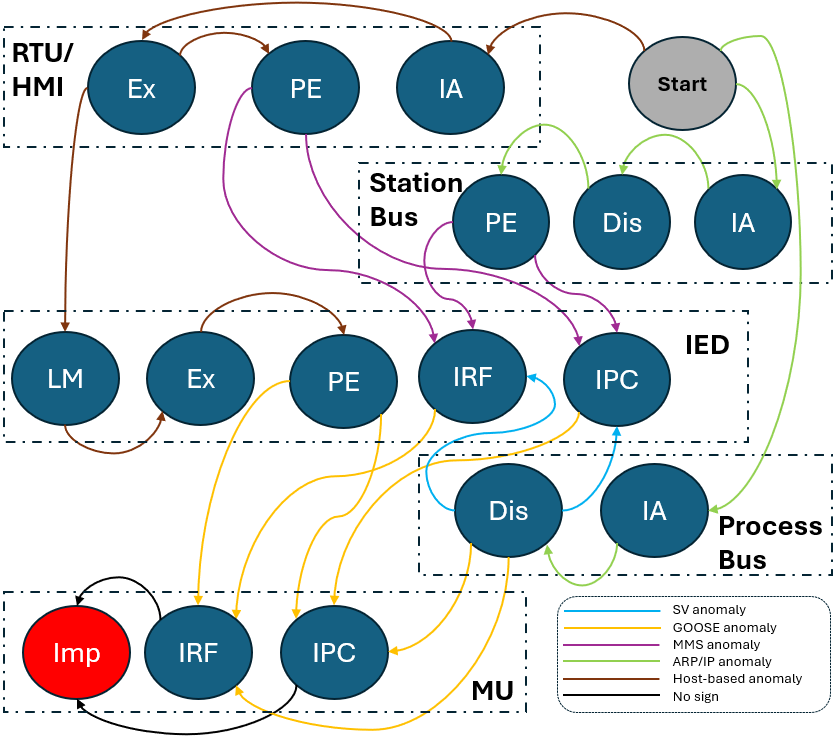}
\caption{Attack graph obtained from the cyber threats modeling.}
\label{fig_1}
\end{figure}

The attack graph shown in Fig. 2 illustrates how the identified attack phases are linked to form end-to-end paths culminating in circuit-breaker (CB) operation disruption. Nodes denote an attack phase at a specific device. The dummy node 'Start' is used to denote the adversary's progression to initial attack phases. Directed edges encode the ATT\&CK techniques that enable progression between attack phases \cite{prob}. These techniques emit anomalies that can be observed by the deployed IDSs. Edge colors indicate the anomaly family (see legend). It should also be noted that there are no anomaly signs for the final phase, as it is at a hard-wired connection. Fig. 2 shows a 1-IED/1-MU example; the attack graph can be scaled to any device count by duplicating the IED/MU subgraphs and preserving inter-type edges.

\section{Hidden Markov Model for SubCASP}

The SubCASP method is designed as a Markov process. This stochastic approach enables the anticipation of potential cyber attack phases in the adversary's broader strategy, thereby enhancing the ability to detect and mitigate them in real-time. In the proposed method, IDS data logs from the substation are received by a trained HMM, as illustrated in Fig. 3.  Then, the current/next attack phases at each time step, and the attack phase sequence at the end of the attack are predicted. The attack phases are modeled as equal-duration discrete time steps, abstracting real-world phase durations to focus on phase ordering and tractable HMM inference.

\begin{figure}[!t]
\centering
\includegraphics[width=0.5\textwidth]{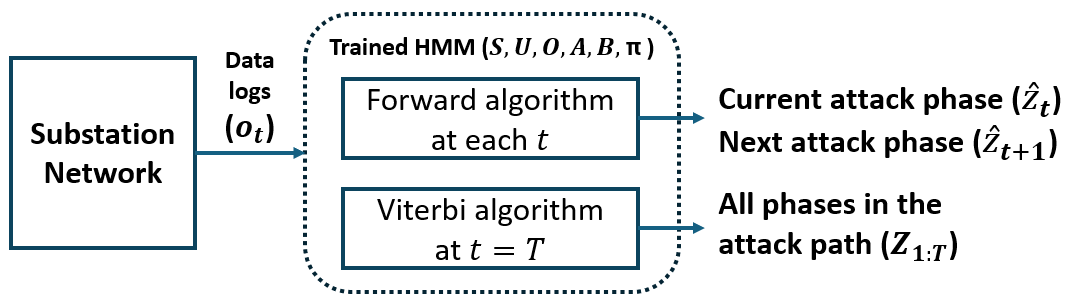}
\caption{Overview of the SubCASP method.}
\label{fig_1}
\end{figure}

In this application, states are the attack phases, which are the nodes in the graph in Fig. 2, represented by a finite set with $N$ states, $S=\{s_{1},s_{2},...,s_{N}\}$. The observations are the data logs generated by both network and host-based IDSs implemented in the system, indicated by the anomaly family in Fig. 2. Set $U$ represents the $M$ number of distinct observations as, $U=\{u_{1},u_{2},...,u_{M}\}$. The observation sequence obtained in real-time operation is given by $O=\{o_{1},o_{2},...,o_{T}\}$, and its length $T$ varies based on the adversary's strategy. At each time step $t$, observation set $o_{t}$ $\epsilon$ $U$ is updated. 

The state transition probability matrix $A$ described the probability of transitioning ($A_{ij}$) from $s_{i}$ to $s_{j}$ for $N$ possible states, given as in (1), where $S_{t}$ is the state at time step $t$.
\begin{equation}
A=\{{\{A_{ij}\} }_{N\times N}|{A_{ij}}=P(S_{t+1}=s_{j}|S_{t}=s_{i})\}, 1\leq i, j\leq N\label{eq}
\end{equation}
The observation or emission probability matrix $B$ is comprised of probability vectors $B_j$. It represents the emission probability of observation $o_t$, given that the attack phase $s_{j}$ occurs at $t$.
\begin{equation}
B=\{\{B_j\}_{N\times M}| B_j(o_t)=P(o_t|S_t=s_{j})\}, 1\leq j \leq N.\label{eq}
\end{equation}
The initial probability distribution vector $\Pi$ given below, specifies how likely the attack gets initiated at the phase $s_{i}$.
\begin{equation}
\Pi=\{\{\pi_i\}_{N\times 1}| \pi_i=P(S_{0}=s_{i})\}, 1\leq i \leq N\label{eq}
\end{equation}

\textbf{\textit{The Forward algorithm (FA)}} is used to estimate the current attack phase and forecast the next phase by propagating the filtered belief through $A$ \cite{FA}. Scaled forward probabilities are used for numerical stability.\\
\textit{Initialization:} The forward probability for each attack phase $s_{j}$ is initially estimated as, $\alpha_1(j)=\pi_j\,B_j(o_1)$.\\
\textit{Recursion:} For each subsequent time step $t$, the forward probability for $s_{j}$ is obtained as in (4). The per-state belief for $s_{j}$ at each $t$ is obtained by (5), while the \textit{argmax} of all the per-state beliefs gives the current attack phase as in (6).

\begin{equation}
\alpha_t(j)=B_j(o_t)\sum_i \alpha_{t-1}(i)\,A_{ij},\quad
\end{equation}
\begin{equation}
b_t(j)=\frac{\alpha_t(j)}{\sum_{k}\alpha_t(k)}.
\end{equation}
\begin{equation}
\hat z_t=\arg\max_j b_t(j).
\end{equation}
Similarly, the next attack phase is predicted by initially obtaining the predictive belief for each state by $q_{t+1}=A^\top b_t$, and then picking the maximum as given below.
\begin{equation}
\hat z_{t+1}=\arg\max_j q_{t+1}(j),\qquad
\end{equation}

\textbf{\textit{The Viterbi algorithm (VA}}) is used to determine the most likely sequence of attack phases at $T$ \cite{viterbi}. The end of the attack sequence at $T$ is identified once FA predicts the final attack phase. The steps involved in this VA application are as follows.\\
\textit{Initialization:} The best score to start in attack phase $s_{j}$, $\delta_1(s_{j})=\pi_j\,B_j(o_1)$. There is no predecessor at the initial time step of the attack $t=1$, so $\psi_1(s_{j})=0$. \\
\textit{Recursion:} To reach $s_{j}$ at time $t$, take the best predecessor $s_{i}$ at $t\!-\!1$ (score $\delta_{t-1}(s_{i})$), transition with probability $A_{ij}$, and emit $o_t$ in $s_{j}$ with probability $B_j(o_t)$. The index of the maximizing predecessor is stored in $\psi_t(s_{j})$, as given below.
\begin{equation}
\delta_t(s_{j})=B_j(o_t)\,\max_{1\le i\le N}\big[\delta_{t-1}(s_{i})\,A_{ij}\big] \label{eq}
\end{equation}
\begin{equation}
\psi_t(s_{j})=\arg\max_{1\le i\le N}\big[\delta_{t-1}(s_{i})\,A_{ij}\big]\label{eq}
\end{equation}
\textit{Termination/backtrack:} The most recent attack phase at $T$ is obtained by (10). Tracing back through $\psi$ reconstructs the most likely sequence $z_{1:T}$, applying the condition $z_{t-1}=\psi_t(z_{t})$.
\begin{equation}
z_{T}=\arg\max_j \delta_T(j) \label{eq}
\end{equation}

\section{Dataset Generation}

A reproducible attack-graph dataset is generated from the ATT\&CK-based graph in Fig. 2. All feasible paths set $P$ =\{$p$\}, are enumerated by depth-first search (DFS) from the 'Start' node to the 'Imp' node. To reflect an adversary’s tendency to pursue the easiest route, paths are weighted by $w(p)$ given in (11). The paths that start externally receive higher mass than internally initiated ones with entry point based weight $w_{entry(p)}$. In addition, shorter paths are favored by a decay function based on path length $L(p)$. The resulting distribution is then used to sample a multiset of paths that forms the training/testing dataset.
\begin{equation}
w(p)=w_{entry(p)}.e^{-\lambda.L(p)}, \lambda > 0 \label{eq}
\end{equation}
Based on the training dataset, the HMM parameters $\Pi, A, B$ are estimated in a supervised manner using maximum likelihood with Laplace smoothing. Disallowed transitions are enforced as structural zeros via an attack-graph mask.

\section{Results and Discussion}

\subsection{SubCASP application for different IDS observability levels}

The performance of the proposed attack-phasing method depends on the information supplied by the IDSs deployed in the substation. Different IDSs provide different levels of observability: some network-IDSs only report anomaly status of a certain communication protocol \cite{mlsdn}, while others additionally identify the target device or even the violated rule \cite{ids}. On the other hand, host-based IDSs range from basic alerts (anomaly/ no-anomaly) to high-level logs that describe the activity. To evaluate the robustness under realistic detection constraints, we consider three IDS-observability levels. 
\begin{itemize}
\item \textbf{Low:} IDSs only report anomaly status (Yes/No).
\item \textbf{Medium:} IDSs report anomaly status plus target device.
\item \textbf{High:} IDSs report anomaly status, target device, and activity type or rule violation.
\end{itemize}

For each case, separate datasets were generated, and HMM parameters were determined. Based on the observability case and the number of devices in the substation, the number of attack phase types ($N$) and the observations ($M$) are varied. For a system with 3 IEDs, $N$=26, and $M$=7, 18, and 29 for the 3 observability levels, low, medium, and high, respectively. 

\textit{(i) Attack Demonstration:}The current phase and next phase prediction obtained by SubCASP using the FA are demonstrated for an attack campaign in the 3 IEDs system. The adversary follows the example attack scenario introduced in Section II, which is initiated at the RTU and ends up as a malicious CB operation. Fig. 4 illustrates the predictions at each $t$, where y-axes show the attack phases arranged in the order of the ATT\&CK matrix.  At low observability levels, SubCASP failed to correctly predict the phases between $t$=3-5. When additional information regarding the target device became available at the medium level, the errors in predictions related to the target IED are corrected (from IED1 to IED3). However, the medium level still produced a false prediction for the \textit{'Impair Process Control'} phase at MU. With the inclusion of high-level observability logs, which specify the exact type of malicious activity, the model accurately predicted the missed phase at MU, along with all 7 phases of the campaign. Furthermore, the next-phase prediction at each time step also improved as the observability level increased from low to high. At $t$=2, the forward belief distribution reflects only one prior transition within the RTU, which provides insufficient evidence to infer the next phase accurately. By $t$=5, although the model correctly anticipates that the next phase will occur at the MU level, the accumulated belief values are not distinctive enough to pinpoint the exact MU phase. Higher observability (more detailed IDS logs) enhances the model’s ability to form sharper belief distributions and, consequently, more accurate current and next-phase predictions.

\textit{(ii) Accuracy Comparison:} The per-step (per-attack phase) prediction accuracies of the current and next-phase are compared with predictions made by VA when predicting the entire attack sequence. The comparison is given in Fig. 5. The results were obtained using five-fold cross-validation over all generated attack paths to prevent dataset bias. The next-phase prediction accuracy using FA remains around 60\% across all IED counts, increasing to above 70\% when the observability level is high. The FA-based current-state detection closely follows the VA results for all observability levels, demonstrating that FA can achieve comparable accuracy in real-time without requiring the entire observation sequence. With a high observability, both FA-current and VA reach above 96\% average per-phase accuracy, confirming the robustness of the proposed HMM parameterization and its scalability with an increasing number of IEDs.

\begin{figure}[!ht]
    \centering
    \includegraphics[width=\linewidth]{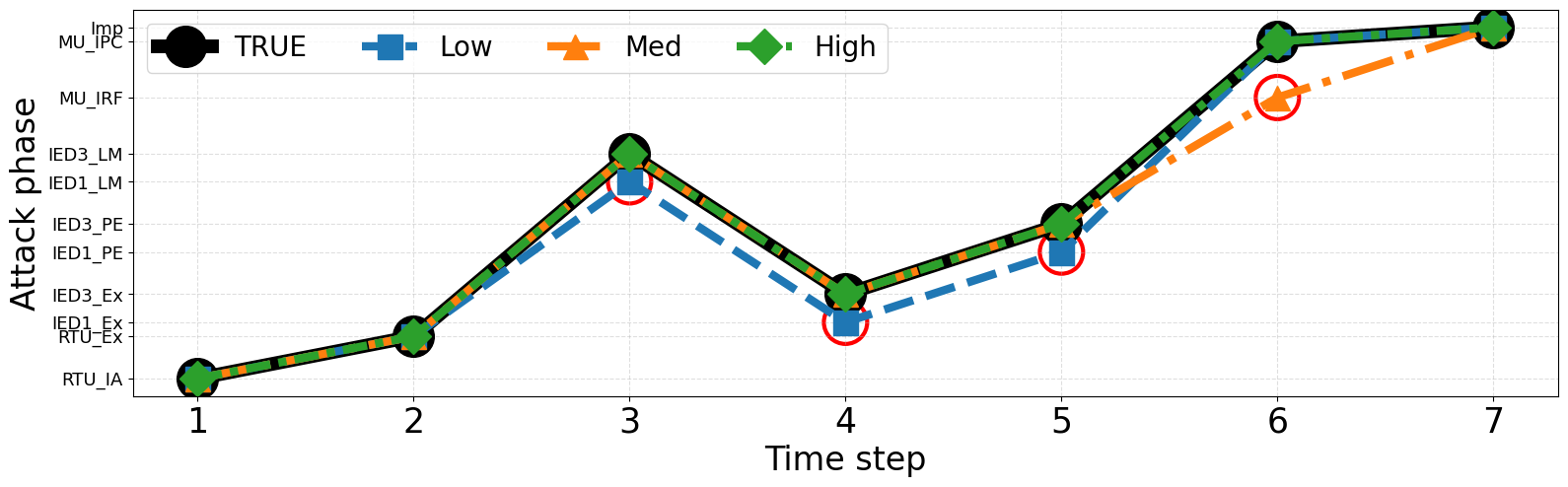}\\[-0.3em]
    {(a)}\\[0.5em]
    \includegraphics[width=\linewidth]{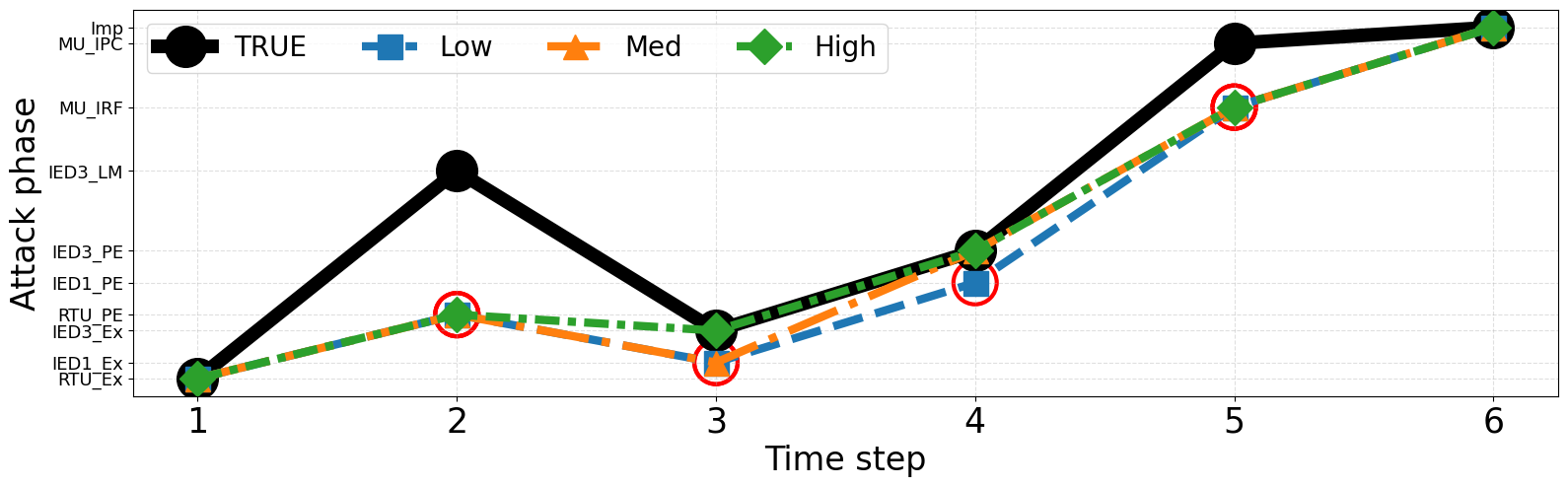}\\[-0.3em]
    {(b)}\\[0.5em]
    \caption{SubCASP predictions at each time step ($t$) for the attack scenario defined in Section II: (a) current attack phase, (b) next attack phase.}
    \label{fig:results}
\end{figure}

\begin{figure}[!t]
\centering
\includegraphics[width=0.5\textwidth]{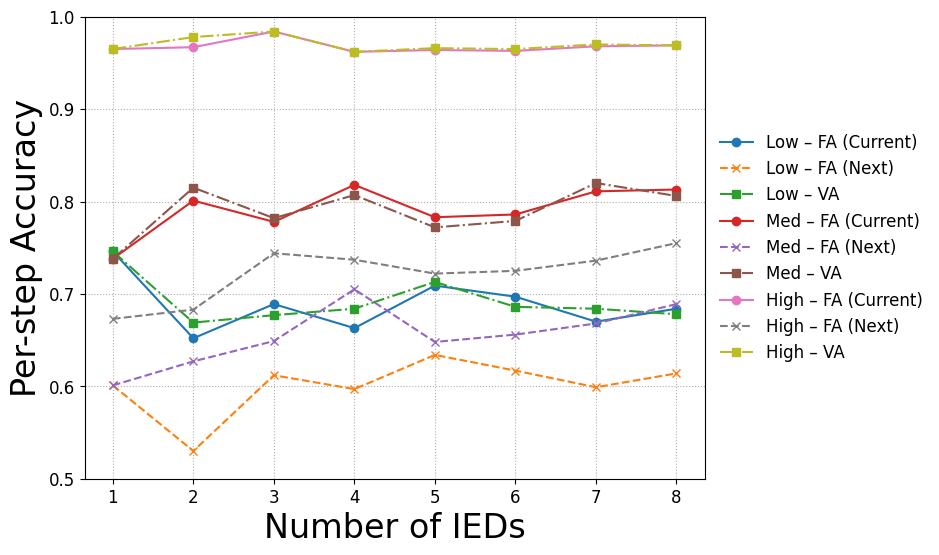}
\caption{Comparison of the accuracy at different IDS observability levels.}
\label{fig_1}
\end{figure}

\subsection{SubCASP testing for missing data logs}

To assess model robustness under missing IDS logs, per-phase detection accuracy was evaluated for varying IED counts using the high-observability IDS case. The comparison between FA for current-state prediction and VA for full-sequence decoding was conducted for 10\%, 20\%, and 30\% missing data (Fig. 6). While both methods yield comparable accuracy with complete logs, VA maintains higher resilience as missing data increases. The maximum deviation between FA and VA increases from 6\% at 10\% to 11\% at 30\%, where VA still achieves over 90\% precision in most IED counts. These results highlight the inclusion of VA in SubCASP, which aids in reconstructing the complete attack sequence and preserving phase-level fidelity even when partial observations are lost.

\begin{figure}[!t]
\centering
\includegraphics[width=0.5\textwidth]{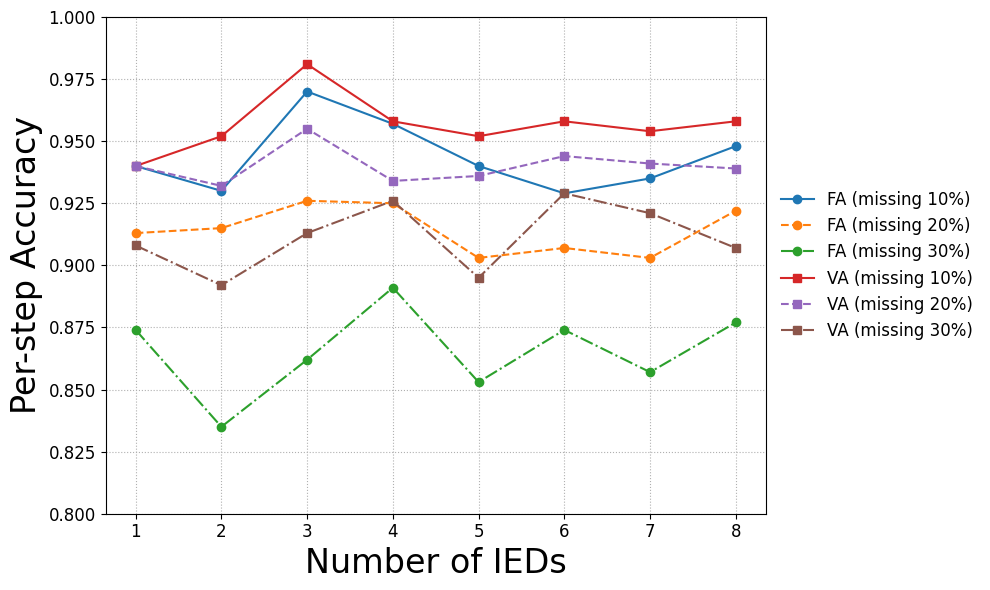}
\caption{Comparison of the accuracy for missing IDS data logs scenarios.}
\label{fig_1}
\end{figure}

\section{Conclusion and Future Work}

This paper proposed the methodology of SubCASP, an HMM-based attack-phasing method that turns heterogeneous IDS logs into phase-aware situational awareness for IEC 61850 digital substations. SubCASP utilizes FA for real-time current/next-phase inference, while VA for retrospective path reconstruction, using an ATT\&CK-informed attack graph. On a reproducible, graph-generated dataset across different IED counts and varying observability levels, FA's current phase prediction closely tracked with VA results, achieving 96\% per-phase prediction accuracy for high IDS observability. The prediction accuracy with VA sustained over 90\% with 30\% missing IDS logs, highlighting the robustness to partial data. The results indicate that reliable attack phase context can be determined by the proposed systematic processing of the IDS logs. For the future work, SubCASP will be modeled for variable attack phase durations. In addition, the scope can be broadened to include impacts other than CB manipulation. The translation of attack phase predictions to actionable mitigation responses will enhance the utility of SubCASP method.

\section*{Acknowledgment}

This research was sponsored by the Director of
Cybersecurity, Energy Security, and Emergency Response,
specifically through the Cybersecurity for .Energy Delivery
Systems program of the U.S. Department of Energy under
contract DE-CR0000021. 

\bibliographystyle{IEEEtran}
\bibliography{references}

\end{document}